\documentclass[preprint]{elsarticle}

\usepackage{lineno,hyperref}
\modulolinenumbers[5]

\journal{Physica B}

\begin{document}

\begin{frontmatter}

\title{Dynamical quadrupole structure factor of frustrated ferromagnetic chain}

\author{Hiroaki Onishi}
\ead{onishi.hiroaki@jaea.go.jp}
\address{Advanced Science Research Center, Japan Atomic Energy Agency, Tokai, Ibaraki 319-1195, Japan}

\begin{abstract}
We investigate the dynamical quadrupole structure factor
of a spin-1/2 $J_{1}$-$J_{2}$ Heisenberg chain
with competing ferromagnetic $J_{1}$ and antiferromagnetic $J_{2}$ in a magnetic field
by exploiting density-matrix renormalization group techniques.
In a field-induced spin nematic regime,
we observe gapless excitations at $q=\pi$
according to quasi-long-range antiferro-quadrupole correlations.
The gapless excitation mode has a quadratic form at the saturation,
while it changes into a linear dispersion as the magnetization decreases.
\end{abstract}


\end{frontmatter}


\section{Introduction}

The spin nematic state,
which is a spin analogue of the nematic liquid crystal,
has attracted much attension as a novel quantum state emerging in magnetic materials
\cite{Andreev1984,Penc-book2011}.
The spin nematic order occurs
when the conventional magnetic dipole order is suppressed
and instead a higher-order quadrupole order appears,
in which spins fluctuate in an axis without its direction along the axis chosen.
In general, the suppression of the magnetic order is caused by
the competition of interactions between spins.
Indeed, several spin models involving competing interactions have been pointed out
to exhibit spin nematic states
\cite{Chubukov1991,Kecke2007,Hikihara2008,Sudan2009,Harada2007,Shannon2006,Tsunetsugu2006,Lauchli2006,Momoi2012}.
In particular,
in a magnetic field,
spins are forced to point to the direction of the magnetic field,
and transverse spin degrees of freedom remain active.
However, the transverse magnetic dipole order is suppressed due to the spin frustration.
Instead of spin operators itself,
the product of spin operators would provide new degrees of freedom
that can possibly order.

As a prototypical model system for the spin nematics,
we focus on a spin-1/2 $J_{1}$-$J_{2}$ Heisenberg chain in a magnetic field
\cite{Chubukov1991,Kecke2007,Hikihara2008,Sudan2009}.
Note that the quadrupole operator is defined on a bond connecting sites
in the spin-1/2 case,
while it can also be defined at single site for $S>1$.
The precise ground-state phase diagram has been obtained theoretically
\cite{Hikihara2008,Sudan2009}.
At high magnetic fields,
the ground state is a spin nematic state,
in which quadropole correlations are quasi-long-ranged
and transverse spin correlations are short-ranged.
Longitudinal spin correlations are also quasi-long-ranged.
The ground state changes to a vector chiral state at low magnetic fields.
To explore the spin nematic state,
a series of edge-sharing copper-oxide chains has been studied
\cite{Masuda2011,Svistov2011,Mourigal2012,Nawa2013,Hase2004,Willenberg2012,Nawa2014}.
However, it is difficult to identify the spin nematic state,
since magnetic probes are usually insensitive to quadrupole correlations,
i.e., four-point spin correlations.

To characterize the spin nematic state from a viewpoint of spin excitations,
dynamical properties such as the NMR relaxation rate
and the dynamical spin structure factor have been studied theoretically
\cite{Sato2009,Sato2011,Onishi2015a,Onishi2015b}.
Recently, it has been pointed out that quadrupole correlations are directly accessible
through ESR measurements
\cite{Furuya2017}.
In this paper,
to gain deep insight into the excitation dynamics in the quadrupole channel,
we investigate the dynamical quadrupole structure factor by exploiting numerical methods.

\section{Model and method}

We consider a spin-1/2 $J_{1}$-$J_{2}$ Heisenberg model
with ferromagnetic $J_{1}<0$
and antiferromagnetic $J_{2}>0$
in a magnetic field $h$
on a one-dimentional chain with $N$ sites,
described by
\begin{equation}
  H =
  J_{1} \sum_{i} \mbox{\boldmath $S$}_{i} \cdot \mbox{\boldmath $S$}_{i+1}
  + J_{2} \sum_{i} \mbox{\boldmath $S$}_{i} \cdot \mbox{\boldmath $S$}_{i+2}
  - h \sum_{i} S_{i}^{z}.
\label{eq: H}
\end{equation}
Throughout the paper,
we take $J_{2}=1$ as the energy unit.
Note that the total magnetization $m=\sum_{i} S_{i}^{z}/N$ is a conserved quantity,
so that it can be used to block-diagonalize the Hamiltonian.
For the calculation of physical quantities at a given $m$,
the magnetic field is chosen to be the midpoint of the magnetization plateau of $m$,
while it is the saturation field for $m=1/2$.

To clarify the excitation dynamics in the quadrupole channel,
we calculate a dynamical quadrupole structure factor,
defined by
\begin{equation}
  Q^{--}(q,\omega) =
  -{\rm Im} \frac{1}{\pi}
  \langle \psi_{\rm G} \vert
  (Q_{q}^{--})^{\dag}
  \frac{1}{\omega+E_{\rm G}-H+{\rm i}\eta}
  Q_{q}^{--}
  \vert \psi_{\rm G} \rangle,
\label{eq: Qmmqw}
\end{equation}
where $Q_{q}^{--}$ is the Fourier transform of $Q_{i}^{--}=S_{i}^{-}S_{i+1}^{-}$,
$\vert \psi_{\rm G} \rangle$ is the ground state,
$E_{\rm G}$ is the ground-state energy,
and $\eta$ is a small broadening factor,
set to 0.1 in the present calculations.

We use density-matrix renormalization group (DMRG) techniques
with open boundary conditions
\cite{White1992,Jeckelmann2002}.
Equation~(\ref{eq: Qmmqw}) is precisely evaluated by targeting multiple states
$\vert \psi_{\rm G} \rangle$,
$Q_{q}^{--}\vert \psi_{\rm G} \rangle$,
and
$[\omega+E_{\rm G}-H+{\rm i}\eta]^{-1}Q_{q}^{--}\vert \psi_{\rm G} \rangle$.
Note that the spectrum at $q$ and $\omega$ is computed
after one run with fixed $q$ and $\omega$,
so that we need to perform many runs
to obtain a full spectrum
in a wide range of the $q$-$\omega$ space.
The Fourier transform of the quadrupole operator
in open boundary conditions is given by
\begin{equation}
  Q_{q}^{--} = \sqrt{\frac{2}{N_{b}+1}}\sum_{i}Q_{i}^{--}\sin(qi),
\end{equation}
where $N_{b}=N-1$ is the number of nearest-neighbor bonds,
and $q=n\pi/(N_{b}+1)$ with integer $n$ $(=1,\cdots,N_{b})$.
We perform DMRG calculations with 64 sites at typical values of $J_{1}$ and $m$,
in which the ground state is in the spin nematic regime at high fields.

\section{Results}

\begin{figure}[t]
\begin{center}
\includegraphics[scale=0.7]{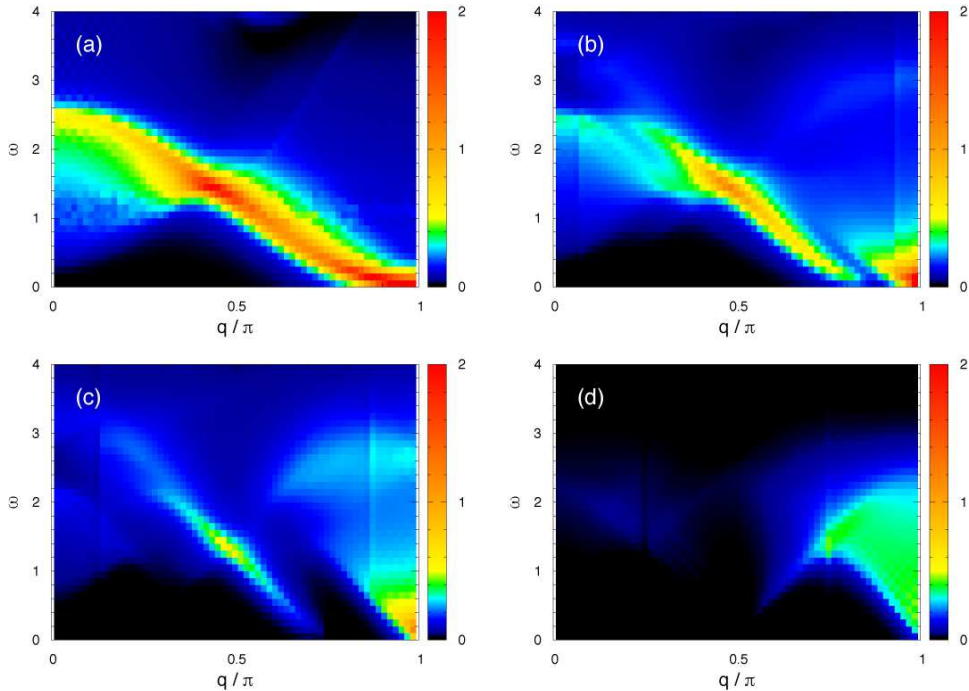}
\end{center}
\caption{
The dynamical quadrupole structure factor $Q^{--}(q,\omega)$
at $J_{1}=-1$ and $J_{2}=1$ for typical values of $m$:
(a) $m=0.5$ (saturation),
(b) $m=0.375$,
(c) $m=0.25$, and
(d) $m=0$,
where $h=1.249$, $1.222$, $1.057$, and $0$, respectively.
The system size is $N=64$.
}
\label{fig1}
\end{figure}

\begin{figure}[t]
\begin{center}
\includegraphics[scale=0.7]{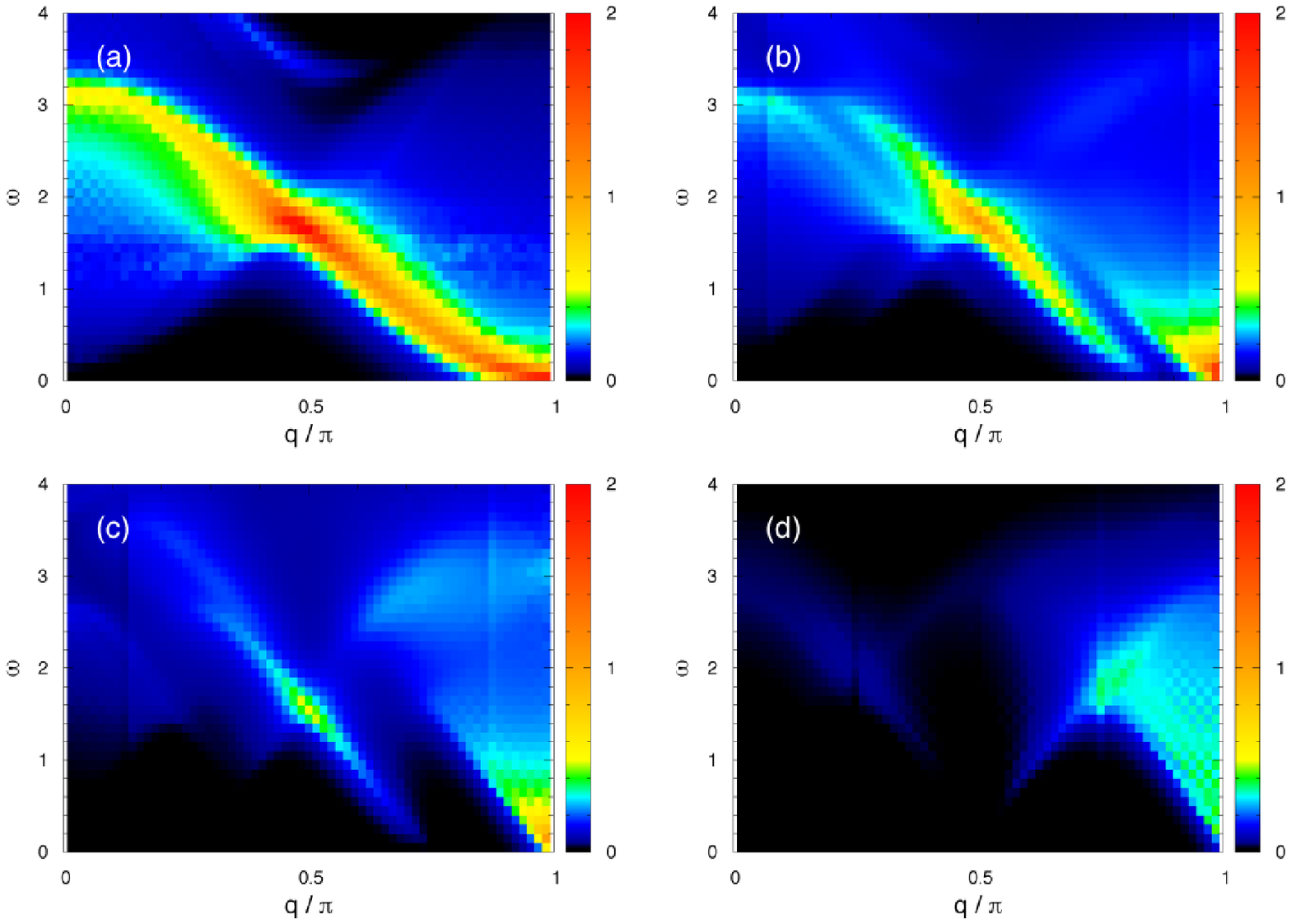}
\end{center}
\caption{
The dynamical quadrupole structure factor $Q^{--}(q,\omega)$
at $J_{1}=-0.5$ and $J_{2}=1$ for typical values of $m$:
(a) $m=0.5$ (saturation),
(b) $m=0.375$,
(c) $m=0.25$, and
(d) $m=0$,
where $h=1.582$, $1.538$, $1.303$, and $0$, respectively.
The system size is $N=64$.
}
\label{fig2}
\end{figure}

\begin{figure}[t]
\begin{center}
\includegraphics[scale=0.7]{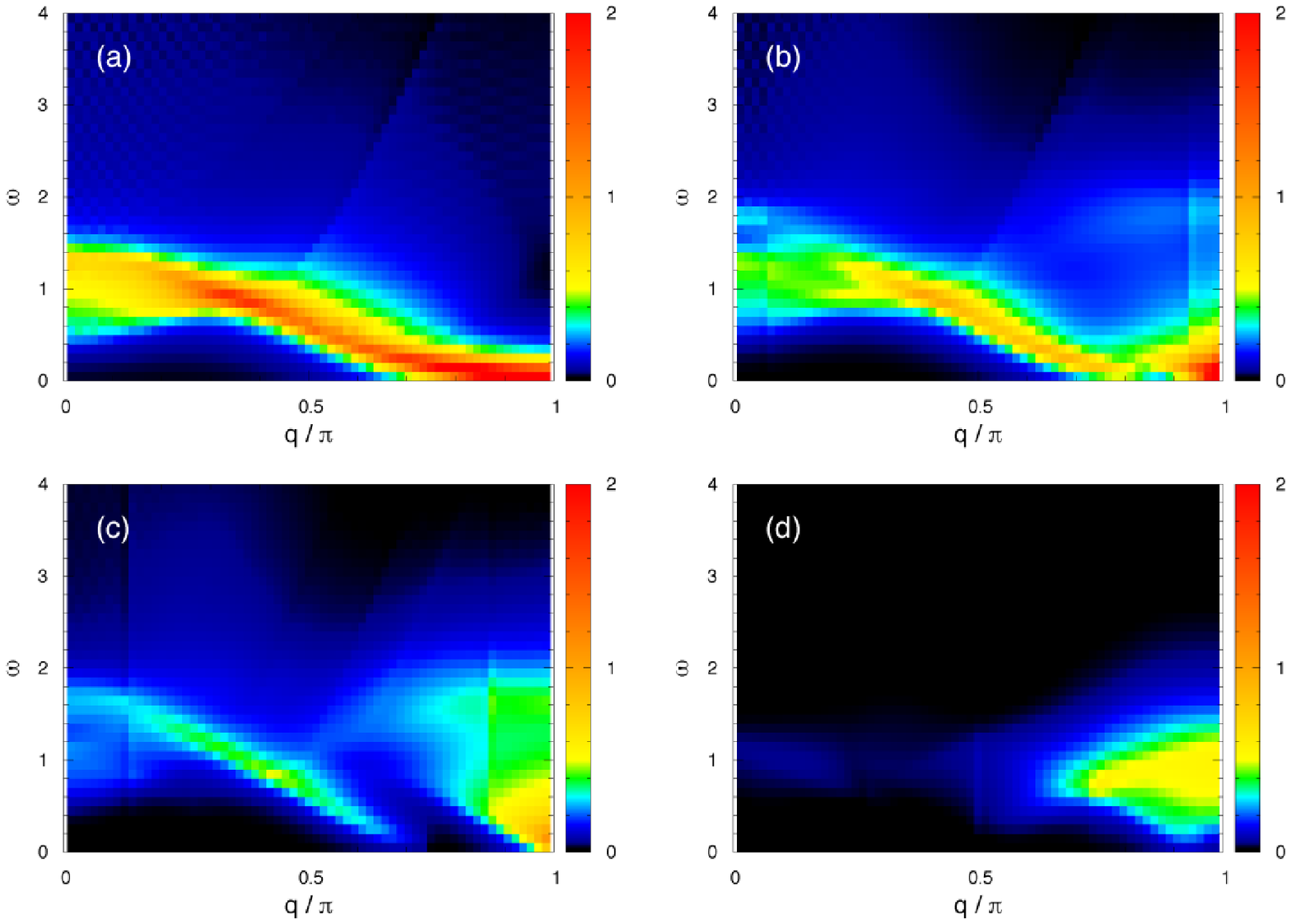}
\end{center}
\caption{
The dynamical quadrupole structure factor $Q^{--}(q,\omega)$
at $J_{1}=-2$ and $J_{2}=1$ for typical values of $m$:
(a) $m=0.5$ (saturation),
(b) $m=0.375$,
(c) $m=0.25$, and
(d) $m=0$,
where $h=0.666$, $0.656$, $0.575$, and $0$, respectively.
The system size is $N=64$.
}
\label{fig3}
\end{figure}

In Fig.~1,
we present intensity plots of $Q^{--}(q,\omega)$
at several values of $m$ for $J_{1}=-1$.
At the saturation $m=0.5$,
we find a dispersive mode extending over the whole range of $q$.
The lowest-energy peak is at $(q_{0},\omega_{0})=(63\pi/64,0.00)$,
indicating gapless quadrupole excitations.
The momentum $63\pi/64$ is the most nearest to $\pi$
among all possible values $n\pi/64$ $(n=1,\cdots,63)$,
suggesting that it converges to $\pi$ in the thermodynamic limit.
Thus the gapless point is at $q=\pi$,
signaling quasi-long-range antiferro-quadrupole correlations.
The dispersion has a quadratic form at low energies,
while it disperses to high energy monotonically
with decreasing $q$.
Comparing peak heights,
the lowest-energy peak at $(q_{0},\omega_{0})$ is most intense.
With decreasing $q$,
the peak height becomes lower,
while it turns to be higher and has a maximum at $(q_{1},\omega_{1})=(28\pi/64,1.40)$,
and again it becomes lower until $q$ reaches the smallest value $\pi/64$.
Continuum excitations are observed for $q<q_{1}$.
The upper bound of the continuum is clearly visible due to the large intensity,
but the lower bound is less distinct since the intensity decreases with approaching there.
The results are consistent with the previous studies of
the energy dispersion for the two-magnon band
\cite{Kecke2007}.

With decreasing $m$,
the excitation mode appears to split into two parts,
as shown in Figs.~1(b) and 1(c).
One represents gapless excitations,
which locates at $q=\pi$ and has large intensity near the gapless point
even when we vary $m$ in the spin nematic regime.
We note that the lowest-energy peak is actually found at
$(q_{0},\omega_{0})=(63\pi/64,0.04)$ for $m=0.375$
and $(63\pi/64,0.08)$ for $m=0.25$.
The finite value of $\omega_{0}$ comes from the finite-size effect.
For instance,
the quadrupole excitation energy and
the width of the magnetization plateau are both finite in
finite-size chains.
$(q_{0},\omega_{0})$ converges to $(\pi,0)$ in the thermodynamic limit
since the gapless point is at $q=\pi$
due to the quasi-long-range antiferro-quadrupole order.
Here, a characteristic feature is that
the gapless excitation mode near $q=\pi$ changes
to a linear dispersion with decreasing $m$.
The other is dispersive around $(q_{1},\omega_{1})$.
Looking at the intensity distribution near $(q_{1},\omega_{1})$,
the spectral weight is quickly suppressed
as we move away from $(q_{1},\omega_{1})$
and the dispersion seems to be disconnected from the gapless excitation mode.
At $m=0$,
where the ground state is not the spin nematic state but a spin singlet state,
we find a sinusoidal dispersion around $q=\pi$,
and the spectral weight is distributed to a continuum above it,
as shown in Fig.~1(d).
The bottom of the dispersion is found at a finite energy,
suggesting that the quadrupole excitation gap opens.
However, the gap is tiny for $J_{1}=-1$
since it is extrapolated to almost zero in the thermodynamic limit.

Let us move on to the investigation at different values of $J_{1}$.
As $|J_{1}|$ decreases with $J_{2}=1$ fixed,
the system approaches decoupled antiferromagnetic chains.
In Fig.~2, we show intensity plots of $Q^{--}(q,\omega)$ for $J_{1}=-0.5$.
We find that the spectrum at each $m$ has a quite similar structure
to that for $J_{1}=-1$.
On the other hand,
as $|J_{1}|$ increases,
the competition between $J_{1}$ and $J_{2}$ interactions becomes significant.
In Fig.~3, we show intensity plots of $Q^{--}(q,\omega)$ for $J_{1}=-2$.
At the saturation,
the bottom of the gapless mode near $q=\pi$ becomes rather shallow and has a flat structure
\cite{Kecke2007}.
The occurrence of a lot of quasi-degenerate states should originate from the spin frustration effect.
With decreasing $m$,
the excitation mode split into two parts in the same way as the case of $J_{1}=-1$.
We find that
the position of $(q_{1},\omega_{1})$ varies with $J_{1}$
such that $q_{1}$ shift toward small momentum and $\omega_{1}$ moves to lower energy
with increasing $|J_{1}|$.
At $m=0$,
the quadrupole excitation gap at $q=\pi$ is more distinct for $J_{1}=-2$ [Fig.~3(d)],
comparing with that for $J_{1}=-1$ [Fig.~1(d)].

\section{Summary}

We have investigated the quadrupole excitation dynamics
of the spin-1/2 one-dimensional $J_{1}$-$J_{2}$ Heisenberg model
in the magnetic field by numerical methods.
In the spin nematic regime,
we have observed gapless excitations at $q=\pi$,
signaling quasi-long-range antiferro-quadrupole correlations,
while the overall spectral structure depends on the magnetization.
The gapless excitation mode changes from
a quadratic form at the saturation into a linear dispersion
with decreasing the magnetization.

\section*{Acknowledgments}

This work was supported by JSPS KAKENHI Grant Number JP16K05494.
Computations were done on the supercomputers at
the Japan Atomic Energy Agency and
the Institute for Solid State Physics, the University of Tokyo.

\section*{References}


\end{document}